\documentclass[aps,pre,twocolumn,superscriptaddress,nofootinbib,10pt]{revtex4-2}

\usepackage[utf8]{inputenc}
\usepackage{tabularx}
\usepackage{multirow}
\usepackage{url}
\usepackage{amsmath}
\usepackage{amssymb}
\usepackage{graphicx}
\usepackage{mathrsfs}
\usepackage{color}
\usepackage{bm}
\usepackage[normalem]{ulem}
\usepackage{float}
\usepackage{lmodern}
\usepackage{soul}
\usepackage[breaklinks,colorlinks,urlcolor=blue,citecolor=blue]{hyperref}

\usepackage{amsmath}
\usepackage{mathrsfs}
\allowdisplaybreaks

\def\simlt{\lower.5ex\hbox{$\; \buildrel < \over \sim \;$}}
\def\simgt{\lower.5ex\hbox{$\; \buildrel > \over \sim \;$}}

\newbox\grsign \setbox\grsign=\hbox{$>$} \newdimen\grdimen \grdimen=\ht\grsign
\newbox\simlessbox \newbox\simgreatbox \newbox\simpropbox
\setbox\simgreatbox=\hbox{\raise.5ex\hbox{$>$}\llap
     {\lower.5ex\hbox{$\sim$}}}\ht1=\grdimen\dp1=0pt
\setbox\simlessbox=\hbox{\raise.5ex\hbox{$<$}\llap
     {\lower.5ex\hbox{$\sim$}}}\ht2=\grdimen\dp2=0pt
\setbox\simpropbox=\hbox{\raise.5ex\hbox{$\propto$}\llap
     {\lower.5ex\hbox{$\sim$}}}\ht2=\grdimen\dp2=0pt
\def\simgt{\mathrel{\copy\simgreatbox}}
\def\simlt{\mathrel{\copy\simlessbox}}

\begin{document}

\title{Absorption of strong electromagnetic waves in magnetized pair plasmas}

\author{Emanuele Sobacchi}
\email{emanuele.sobacchi@gssi.it}
\affiliation{Gran Sasso Science Institute, viale F.~Crispi 7, L’Aquila, 67100, Italy}
\affiliation{INFN -- Laboratori Nazionali del Gran Sasso, via G.~Acitelli 22, Assergi, 67100, Italy}

\begin{abstract}
We discuss synchrotron absorption of a short electromagnetic pulse that propagates in a cold magnetized pair plasma. We show that the pulse can be absorbed when $\omega_{\rm B}/a_0< \omega< a_0\omega_{\rm B}$, where $a_0>1$ is the strength parameter of the pulse, and $\omega$ and $\omega_{\rm B}$ respectively are the frequency of the wave and the cyclotron frequency in the background magnetic field (all quantities are defined in the reference frame where the particles are at rest before being illuminated by the pulse). The condition $\omega_{\rm B}/a_0< \omega< a_0\omega_{\rm B}$ is essentially a generalization of the cyclotron resonance to strong electromagnetic pulses with $a_0>1$. When $\omega_{\rm B}/a_0< \omega< a_0\omega_{\rm B}$, the propagation of electromagnetic waves in a plasma can be very different with respect to the propagation in vacuum because the wave equation is strongly non-linear. Then it is unclear whether the particles are heated stochastically due to synchrotron absorption, as found by studying the motion of a test particle in the field of a vacuum electromagnetic wave. We discuss implications of our results for constraining emission models of fast radio bursts.
\end{abstract}

\maketitle

\section{Introduction}

Fast radio bursts (FRBs) are bright extragalactic radio transients of millisecond duration \citep[][]{CordesChatterjee2019, Petroff+2019, Petroff+2022}. FRBs are likely produced by magnetars \citep[][]{Bochenek+2020, Chime2020}, which are strongly magnetized neutron stars surrounded by a pair plasma \citep[][]{KaspiBeloborodov2017}. The huge radio luminosity of FRBs (isotropic equivalent $L_{\rm iso}\sim 10^{42}{\rm\; erg\; s^{-1}}$ in the GHz band) implies that electrons oscillate with relativistic velocities in the field of the radio wave even at considerable distances from the magnetar, $R\sim 10^{13}{\rm\; cm}$ \citep[][]{LuanGoldreich2014}.

The propagation of strong electromagnetic waves in magnetized pair plasmas is crucial for the physics of FRBs \citep[][]{Lyubarsky2021, Zhang2023}. For example, FRB emission models could be constrained on the basis of whether the radio wave can escape the source \citep[][]{Beloborodov2021, Qu+2022, GolbraikhLyubarsky2023, Beloborodov2024, Lyutikov2024, Sobacchi+2024b, Beloborodov2025}. In the following, we consider linearly polarized waves where the wave electric field and the wave vector are perpendicular to the background magnetic field. The emission of these modes is predicted by several FRB models, including the shock maser model \citep[][]{Lyubarsky2014, Beloborodov2017, Metzger+2019, Beloborodov2020, Sironi+2021, Iwamoto+2024, VanthieghemLevinson2025} and the reconnection model \citep[][]{Lyubarsky2020, Mahlmann+2022}.

The wave-plasma interaction has been investigated by studying the motion of a test particle in the field of a vacuum wave. Lyubarsky \citep[][]{Lyubarsky2018} considered an infinite plane wave and worked in the ``fluid frame'' where the average electric field and the average velocity vanish. In this frame, particles of Lorentz factor $\tilde{\gamma}$ gain energy through synchrotron absorption when $\tilde{\omega}_{\rm B}/\tilde{\gamma}< \tilde{\omega}< \tilde{\gamma}^2 \tilde{\omega}_{\rm B}/a_0^3$, where $\tilde{\omega}$ is the wave frequency, $\tilde{\omega}_{\rm B}$ is the cyclotron frequency in the background magnetic field, and $a_0>1$ is the wave strength parameter ($a_0$ is defined as the peak transverse component of the electron four-velocity in units of the speed of light, which is a relativistic invariant). The tilde indicates the quantities measured in the fluid frame.

Beloborodov \citep[][]{Beloborodov2022} numerically simulated the motion of a test particle illuminated by a pulse of finite duration. They worked in the ``lab frame'' where the fluid is at rest before being illuminated and assumed that the average velocity vanishes also when the fluid is illuminated. They showed that particles are heated stochastically when $\omega_{\rm B}/a_0< \omega< \omega_{\rm B}$. In a recent preprint \citep[][]{Beloborodov2025}, Beloborodov argues that particles can be heated stochastically by high-frequency waves (up to $\omega\sim a_0^2\omega_{\rm B}$) when the bulk Lorentz factor of the illuminated fluid is $\gamma_{\rm bulk}\sim a_0$ in the lab frame.\footnote{Beloborodov initially assumed that bulk acceleration and compression of magnetar winds are negligible during the interaction with FRBs (see Eqs.~11 and 29 of ref.~\citep[][]{Beloborodov2021}). After we showed that bulk acceleration and compression can instead be very large \citep[][]{Sobacchi+2024b}, Beloborodov revised their calculations in ref.~\citep[][]{Beloborodov2025}.}

In our previous work \citep[][]{Sobacchi+2024b}, we studied the propagation of a pulse of finite duration following a different approach. We solved the entire set of Maxwell’s equations, coupled with a two-fluid model of the pair plasma. We assumed that the plasma is strongly magnetized (i.e.~$\omega_{\rm B}>\omega_{\rm P}$, where $\omega_{\rm P}$ is the plasma frequency). In the lab frame, we identified three regimes.\footnote{The composition of the plasma is very important. In unmagnetized electron-proton plasmas, the current is a linear function of the wave field when $a_0<1$, or alternatively when $a_0>1$ and $a_0\omega_{\rm P}\tau_0<1$, where $\tau_0$ is the duration of the electromagnetic pulse in the lab frame \citep[][]{Sprangle+1990b, Sobacchi+2024a}. In unmagnetized pair plasmas, the current is linear when $\omega>a_0\omega_{\rm P}$ \citep[][]{Sobacchi+2024a}.} (i) When $\omega<\omega_{\rm B}/a_0$, the illuminated fluid is non-relativistic (i.e.~$\gamma_{\rm bulk}\sim 1$) because the wave field is smaller than the background magnetic field. The plasma current is a linear function of the wave field. This regime was already well studied \citep[][]{Melrose1997}. (ii) When $\omega>a_0\omega_{\rm B}$, the bulk Lorentz factor of the fluid is $\gamma_{\rm bulk}\sim a_0$, as in the absence of a background magnetic field  \citep[][]{GunnOstriker1971, LandauLifshitz1975, Sobacchi+2024a}, and the current is linear. (iii) When $\omega_{\rm B}/a_0<\omega<a_0\omega_{\rm B}$, the bulk Lorentz factor of the fluid is relativistic and the current is non-linear (we could not exactly determine the bulk Lorentz factor and the current). We demonstrated that the pulse is not absorbed in the first or second regimes and speculated that it could be absorbed in the third regime. However, we did not identify a specific absorption mechanism.

The results of these studies, which considered different reference frames and adopted different assumptions, are not clearly compatible with each other. In this paper, we reconsider the propagation of a strong electromagnetic pulse in a magnetized pair plasma where $\omega_{\rm B}>\omega_{\rm P}$. We show the following. (i) Stochastic heating of particles described by ref.~\citep[][]{Beloborodov2022} is, in fact, caused by synchrotron absorption, which was previously studied by ref.~\citep[][]{Lyubarsky2018}. (ii) A strong pulse is not absorbed when $\omega<\omega_{\rm B}/a_0$ or $\omega>a_0\omega_{\rm B}$, while it is absorbed when $\omega_{\rm B}/a_0<\omega<a_0\omega_{\rm B}$ (here, all quantities are defined in the lab frame). These results are consistent with our previous study \citep[][]{Sobacchi+2024b} and inconsistent with ref.~\citep[][]{Beloborodov2025}, where the author argues that the pulse can be absorbed when $\omega< a_0^2\omega_{\rm B}$. (iii) Studies that consider the motion of a test particle in a vacuum wave have limited applicability when $\omega_{\rm B}/a_0<\omega<a_0\omega_{\rm B}$. Then it is unclear whether the particles can be heated stochastically due to synchrotron absorption.

The paper is organized as follows. In Sec.~\ref{sec:absorption} we discuss synchrotron absorption of an infinite vacuum plane wave. We show that stochastic heating is equivalent to synchrotron absorption. In Sec.~\ref{sec:pulse} we discuss the absorption of a pulse of finite duration. Since the amplitude of the pulse is not constant, it is important to distinguish the local value of the strength parameter, $a$, and its maximum value, $a_0$. In Sec.~\ref{sec:discussion} we discuss the implications of our results for FRB emission models. Our notation is summarized in Table \ref{table:param}.

\section{Equivalence of synchrotron absorption and stochastic heating}
\label{sec:absorption}

In the following, we outline the theory of synchrotron absorption of strong electromagnetic waves developed by Lyubarsky \citep[][]{Lyubarsky2018}. We consider the motion of a test particle of Lorentz factor $\tilde{\gamma}$ in the field of a monochromatic vacuum electromagnetic plane wave. We work in the ``fluid frame'' where the average electric field vanishes. In the following, the tilde indicates the quantities measured in the fluid frame. We consider an infinite plane wave whose peak electromagnetic fields are $\tilde{E}_{\rm w}=\tilde{B}_{\rm w}$. The frequency of the wave is $\tilde{\omega}$. There is a uniform background magnetic field, $\tilde{B}_{\rm bg}$, aligned with the magnetic field of the wave. The wave strength parameter is\footnote{The amplitude of an infinite plane wave (averaged over several wavelengths) is constant. Then, the local value of the strength parameter, $a$, is equal to its maximum value, $a_0$.} $a=e\tilde{E}_{\rm w}/\tilde{\omega}mc$, and the cyclotron frequency is $\tilde{\omega}_{\rm B}=e\tilde{B}_{\rm bg}/mc$, where $e$ and $m$ are the charge and mass of the electron, and $c$ is the speed of light.

We consider the regime in which the wave frequency, $\tilde{\omega}$, is higher than the particle rotation frequency, $\tilde{\omega}_{\rm L}\sim \tilde{\omega}_{\rm B}/\tilde{\gamma}$, namely
\begin{equation}
\label{eq:omegamin}
\tilde{\omega} > \frac{\tilde{\omega}_{\rm B}}{\tilde{\gamma}} \;.
\end{equation}
The trajectory of the particles can be decomposed into fast oscillations on the timescale $1/\tilde{\omega}$, superimposed on a slow Larmor rotation on the timescale $1/\tilde{\omega}_{\rm L}$. The Lorentz factor of the guiding center is
\begin{equation}
\label{eq:gammagc}
\tilde{\gamma}_{\rm gc}\sim \frac{\tilde{\gamma}}{\max\left[1,a\right]} \;,
\end{equation}
which is equivalent to Eq.~(30) of ref.~\citep[][]{Lyubarsky2018}. This expression can be interpreted as follows. When the wave is weak (i.e.~$a<1$), the fast oscillations are non-relativistic, and the Lorentz factor of the guiding center is equal to $\tilde{\gamma}$. When the wave is strong  (i.e.~$a>1$), the fast oscillations are relativistic. The particle acquires an effective mass $am$, and the Lorentz factor of the guiding center is $\tilde{\gamma}/a$. Synchrotron absorption occurs at high harmonics of the rotation frequency. The characteristic frequency is $\tilde{\gamma}_{\rm gc}^3 \tilde{\omega}_{\rm L}$ \citep[][]{Lyubarsky2018}. When the wave is weak (i.e.~$a<1$), the characteristic frequency is $\tilde{\gamma}^3 \tilde{\omega}_{\rm L}\sim \tilde{\gamma}^2 \tilde{\omega}_{\rm B}$, which is a classical result of synchrotron radiation theory \citep[][]{LandauLifshitz1975}. Above the characteristic frequency, absorption is suppressed. The condition $\tilde{\omega}<\tilde{\gamma}_{\rm gc}^3 \tilde{\omega}_{\rm L}$ can be presented as
\begin{equation}
\label{eq:omegamax}
\tilde{\omega} < \frac{\tilde{\gamma}^2\tilde{\omega}_{\rm B}}{\max\left[1,a^3\right]}\;,
\end{equation}
which is equivalent to Eq.~(63) of ref.~\citep[][]{Lyubarsky2018}. When Eqs.~\eqref{eq:omegamin} and \eqref{eq:omegamax} are satisfied, the particle absorbs energy from the wave. The mean energy gain per rotation time is given by Eq.~(59) of ref.~\citep[][]{Lyubarsky2018}, which can be presented as
\begin{equation}
\langle\Delta\tilde{\gamma}\rangle \sim a^2 \tilde{\gamma} \left(\frac{\tilde{\omega}}{\tilde{\gamma}^2\tilde{\omega}_{\rm B}}\right)^{2/3} \;.
\end{equation}
We can present an equation for the evolution of $\tilde{\gamma}$ in the form ${\rm d}\tilde{\gamma}/{\rm d}\tilde{t}\sim\tilde{\omega}_{\rm L}\langle\Delta\tilde{\gamma}\rangle$, or equivalently
\begin{equation}
\label{eq:gammaevo}
\frac{{\rm d}\tilde{\gamma}}{{\rm d}\tilde{t}} \sim a^2 \tilde{\omega}_{\rm B} \left(\frac{\tilde{\omega}}{\tilde{\gamma}^2\tilde{\omega}_{\rm B}}\right)^{2/3} \;.
\end{equation}

Eq.~\eqref{eq:gammaevo} is the same as Eq.~(108) of ref.~\citep[][]{Beloborodov2024}. Beloborodov \citep[][]{Beloborodov2022, Beloborodov2024} does not recognize that the stochastic heating of particles described by Eq.~\eqref{eq:gammaevo} is, in fact, caused by synchrotron absorption. In a recent preprint, Beloborodov incorrectly assumes that Eq.~\eqref{eq:gammaevo} describes the evolution of $\tilde{\gamma}$ when $\tilde{\omega}_{\rm B}/a<\tilde{\omega}<\tilde{\omega}_{\rm B}$ (see Eq.~10 of ref.~\citep[][]{Beloborodov2025}). The correct validity range of Eq.~\eqref{eq:gammaevo} is given instead by Eqs.~\eqref{eq:omegamin} and \eqref{eq:omegamax}. As we show in Sec.~\ref{sec:pulse}, identifying the correct range of validity is very important for the absorption of short electromagnetic pulses.

\begin{table}
\centering
\begin{tabular}{cc}
symbol & physical quantity\\
\hline
$\tilde{\omega}\sim\omega/\gamma_{\rm bulk}$ & frequency of the electromagnetic pulse \\
$a_0$ & maximum strength parameter of the pulse \\
$a$ & local strength parameter of the pulse \\
$\gamma_{\rm bulk}$ & bulk Lorentz factor of the illuminated fluid\\
$\tilde{\gamma}$ & particle Lorentz factor \\
$\tilde{\gamma}_{\rm gc}$ & Lorentz factor of the guiding center \\
$\tilde{\omega}_{\rm B}\sim\gamma_{\rm bulk}\omega_{\rm B}$ & cyclotron frequency
\end{tabular}
\caption{\label{table:param} Summary of our notation. Quantities without the tilde are measured in the ``lab frame'' where the fluid is at rest before being illuminated by the pulse. Quantities with the tilde are measured in the ``fluid frame'' where the bulk velocity of the illuminated fluid vanishes. The strength parameter is a relativistic invariant.}
\end{table}

\section{Synchrotron absorption of short electromagnetic pulses}
\label{sec:pulse}

In the following, we consider the motion of a test particle in the field of a vacuum electromagnetic pulse, whose amplitude in the fluid frame varies on long spatial scales compared to $c/\tilde{\omega}_{\rm L}$. Since the amplitude of the pulse is not constant, we distinguish the local value of the strength parameter, $a$, and its maximum value, $a_0$. We assume that the wave frequency is higher than the particle rotation frequency (i.e.~$\tilde{\omega}>\tilde{\omega}_{\rm L}$).

Before synchrotron absorption is activated, in the fluid frame the particle Lorentz factor is determined by the fast oscillations and given by $\tilde{\gamma}\sim\max[1,a]$. As shown by Eq.~\eqref{eq:gammagc}, the motion of the guiding center is non-relativistic (i.e.~$\tilde{\gamma}_{\rm gc}\sim 1$). When $\tilde{\gamma}\sim\max[1,a]$, Eqs.~\eqref{eq:omegamin} and \eqref{eq:omegamax} can be satisfied simultaneously only for
\begin{equation}
\label{eq:resonance}
\tilde{\omega} \sim \frac{\tilde{\omega}_{\rm B}}{\max\left[1,a\right]}\;.
\end{equation}
When the wave is weak (i.e.~$a<1$), Eq.~\eqref{eq:resonance} is the cyclotron resonance condition. When the wave is strong (i.e.~$a>1$), Eq.~\eqref{eq:resonance} is the cyclotron resonance condition for a particle of effective mass $am$.

It is important to express the resonance condition in terms of the physical parameters evaluated in the ``lab frame'' where the fluid is at rest before being illuminated by the pulse. In the lab frame, the frequency of the wave is $\omega$. Outside the pulse, the background magnetic field is $B_{\rm bg,0}$, and there is no electric field. We define the cyclotron frequency in the lab frame as $\omega_{\rm B}=eB_{\rm bg,0}/mc$. Inside the pulse, the background magnetic field, $B_{\rm bg}$, can be different from $B_{\rm bg,0}$, and there could be a non-zero average electric field, $E_{\rm bg}$. The velocity of the fluid frame with respect to the lab frame is given by $v_{\rm bulk}/c=E_{\rm bg}/B_{\rm bg}$. The corresponding Lorentz factor is $\gamma_{\rm bulk}=(1-v_{\rm bulk}^2/c^2)^{-1/2}$.

Since the illuminated fluid moves in the same direction as the pulse, the frequency of the wave in the fluid frame is redshifted. Then, we have
\begin{equation}
\label{eq:rel1}
\tilde{\omega} \sim \frac{\omega}{\gamma_{\rm bulk}}\;.
\end{equation}
In the lab frame, the thickness of a fluid shell is compressed by a factor of $(1-v_{\rm bulk}/c)\sim\gamma_{\rm bulk}^{-2}$ within the pulse. We assume that the background magnetic field is frozen in the fluid. Since the background field is perpendicular to the direction of propagation of the pulse, its strength is amplified by a factor of $\gamma_{\rm bulk}^2$.
Then, we have $B_{\rm bg}\sim\gamma_{\rm bulk}^2B_{\rm bg,0}$ (see also Eqs.~8 and 9 of ref.~\citep[][]{Beloborodov2021}). The background magnetic field in the fluid frame is given by $\tilde{B}_{\rm bg}=B_{\rm bg}/\gamma_{\rm bulk}\sim\gamma_{\rm bulk}B_{\rm bg,0}$, which implies
\begin{equation}
\label{eq:rel2}
\tilde{\omega}_{\rm B}\sim\gamma_{\rm bulk}\omega_{\rm B} \;.
\end{equation}
Substituting Eqs.~\eqref{eq:rel1} and \eqref{eq:rel2} into Eq.~\eqref{eq:resonance}, we can present the resonance condition in the lab frame as
\begin{equation}
\label{eq:resonancelab}
\omega\sim \gamma_{\rm bulk}^2\;\frac{\omega_{\rm B}}{\max\left[1,a\right]} \;.
\end{equation}

Synchrotron absorption (equivalently, stochastic heating) is activated if Eq.~\eqref{eq:resonancelab} is satisfied.\footnote{After activation of synchrotron absorption, the Lorentz factor grows according to Eq.~\eqref{eq:gammaevo}. Since the range of frequencies given by Eqs.~\eqref{eq:omegamin} and \eqref{eq:omegamax} becomes very broad when $\tilde{\gamma}\gg\max[1,a]$, synchrotron absorption may continue even if the particle moves to a part of the pulse where Eq.~\eqref{eq:resonancelab} is not satisfied.} The solution of Eq.~\eqref{eq:resonancelab} is not trivial because $\gamma_{\rm bulk}$ could depend on $a$. In Secs.~\ref{sec:pulsenobulk} and \ref{sec:pulsebulk} we consider, respectively, the scenario in which the fluid does not accelerate when illuminated by the pulse and the scenario in which the fluid accelerates at the maximal rate, as in the absence of a background magnetic field. Then, in Sec.~\ref{sec:wave} we use the self-consistent expression of $\gamma_{\rm bulk}$ calculated in our previous work \citep[][]{Sobacchi+2024b}.

\subsection{No bulk acceleration of the fluid}
\label{sec:pulsenobulk}

We consider the scenario in which the illuminated fluid does not accelerate (i.e.~$\gamma_{\rm bulk}=1$). This scenario was considered by Beloborodov \citep{Beloborodov2022}, who numerically simulated the motion of a test particle illuminated by the pulse. When $a<1$, Eq.~\eqref{eq:resonancelab} gives $\omega\sim\omega_{\rm B}$. The wave can be absorbed when its frequency is equal to the cyclotron frequency. When $a>1$, Eq.~\eqref{eq:resonancelab} gives
\begin{equation}
\label{eq:resonancenobulk}
\omega\sim\frac{\omega_{\rm B}}{a} \;.
\end{equation}
Synchrotron absorption is activated when the local value of the strength parameter is $a\sim\omega_{\rm B}/\omega$. The consistency condition $1<a<a_0$ gives
\begin{equation}
\label{eq:rangenobulk}
\frac{\omega_{\rm B}}{a_0} < \omega < \omega_{\rm B}\;.
\end{equation}
Eq.~\eqref{eq:rangenobulk} gives the parameter space region where the pulse could be absorbed if the fluid does not accelerate. Our result is consistent with the numerical simulations presented in ref.~\citep{Beloborodov2022} (see their Eq.~1).

\subsection{Maximal bulk acceleration of the fluid}
\label{sec:pulsebulk}

We consider the scenario in which the illuminated fluid accelerates as in the absence of a background magnetic field. In this scenario, in the lab frame, the Lorentz factor of the particles is $\gamma=1+a^2/2$, and the four-velocity component along the direction of propagation of the wave is $u_\parallel/c=a^2/2$ \citep[][]{GunnOstriker1971, LandauLifshitz1975}. The Lorentz factor of the fluid frame (where the particles make a closed trajectory) with respect to the lab frame is
\begin{equation}
\gamma_{\rm bulk}\sim\max[1,a]\;.
\end{equation}
When $a>1$, Eq.~\eqref{eq:resonancelab} gives
\begin{equation}
\label{eq:resonancebulk}
\omega\sim a\omega_{\rm B} \;.
\end{equation}
Synchrotron absorption is activated when the local value of the strength parameter is $a\sim\omega/\omega_{\rm B}$. The consistency condition $1<a<a_0$ gives
\begin{equation}
\label{eq:rangebulk}
\omega_{\rm B} < \omega < a_0\omega_{\rm B}\;.
\end{equation}
Eq.~\eqref{eq:rangebulk} gives the parameter space region where the pulse could be absorbed if the fluid accelerates at the maximal rate, as in the absence of a background magnetic field.

Our result is inconsistent with a recent preprint by Beloborodov \citep[][]{Beloborodov2025}, who argues that the pulse could be absorbed when $\omega < a_0^2\omega_{\rm B}$. The reason for the discrepancy is the following. Beloborodov assumes that Eq.~\eqref{eq:gammaevo} describes the evolution of $\tilde{\gamma}$ when $\tilde{\omega}_{\rm B}/a<\tilde{\omega}<\tilde{\omega}_{\rm B}$ (see Eq.~10 of ref.~\citep[][]{Beloborodov2025}). Taking into account that $\tilde{\omega}\sim\omega/a_0$ and $\tilde{\omega}_{\rm B}\sim a_0\omega_{\rm B}$ when $\gamma_{\rm bulk}\sim a_0$ (see Eqs.~\ref{eq:rel1} and \ref{eq:rel2}), the condition $\tilde{\omega}<\tilde{\omega}_{\rm B}$ would give $\omega < a_0^2\omega_{\rm B}$. However, as discussed above, Eq.~\eqref{eq:gammaevo} describes the evolution of $\tilde{\gamma}$ when $\tilde{\omega}_{\rm B}/\tilde{\gamma}<\tilde{\omega}<\tilde{\gamma}^2\tilde{\omega}_{\rm B}/a^3$ (see Eqs.~\ref{eq:omegamin} and \ref{eq:omegamax}), and synchrotron absorption is activated when $\tilde{\omega}\sim\tilde{\omega}_{\rm B}/a$ (see Eq.~\ref{eq:resonance}). Eq.~(10) of ref.~\citep[][]{Beloborodov2025} follows from an incorrect interpretation of the simulations of ref.~\citep[][]{Beloborodov2022}. As we discussed in Sec.~\ref{sec:pulsenobulk}, these simulations show that the pulse could be absorbed when $\omega_{\rm B}/a_0<\omega<\omega_{\rm B}$ if $\gamma_{\rm bulk}=1$.

\subsection{Self-consistent expression of the bulk Lorentz factor of the fluid}
\label{sec:wave}

In our previous work \citep[][]{Sobacchi+2024b}, we studied the wave-plasma interaction by solving the entire set of Maxwell's equations, coupled with a two-fluid model of the magnetized pair plasma. The wave does not propagate as in vacuum, and the feedback of the plasma current on the wave is taken into account. The bulk Lorentz factor of the illuminated fluid is given by \citep[][]{Sobacchi+2024b}
\begin{equation}
\label{eq:gammabulk}
\gamma_{\rm bulk} \sim
\begin{cases}
1 & \;\;{\rm when}\;\; a<1 \\
1 & \;\;{\rm when}\;\; a>1\;\;{\rm and}\;\; \omega<\omega_{\rm B}/a \\
a & \;\;{\rm when}\;\; a>1\;\;{\rm and}\;\; \omega>a\omega_{\rm B}
\end{cases}
\;.
\end{equation}
We could not calculate $\gamma_{\rm bulk}$ self-consistently when $a>1$ and $\omega_{\rm B}/a<\omega<a\omega_{\rm B}$. Nevertheless, Eqs.~\eqref{eq:resonancelab} and \eqref{eq:gammabulk} are sufficient to determine whether synchrotron absorption is activated. There are four cases.
\begin{enumerate}
\item When $\omega<\omega_{\rm B}/a_0$, we have $\gamma_{\rm bulk}\sim 1$ in the whole pulse. Absorption is not activated.
\item When $\omega_{\rm B}/a_0< \omega< \omega_{\rm B}$, we have $\gamma_{\rm bulk}\sim 1$ in the leading part of the pulse, where $a<\omega_{\rm B}/\omega$. Absorption is activated when $a\sim\omega_{\rm B}/\omega$.
\item When $\omega_{\rm B}<\omega<a_0\omega_{\rm B}$, we have $\gamma_{\rm bulk}\sim \max[1,a]$ in the leading part of the pulse, where $a<\omega / \omega_{\rm B}$. Absorption is activated when $a\sim\omega / \omega_{\rm B}$. 
\item When $\omega>a_0\omega_{\rm B}$, we have $\gamma_{\rm bulk}\sim \max[1,a]$ in the whole pulse. Absorption is not activated.
\end{enumerate}
We conclude that synchrotron absorption is activated when
\begin{equation}
\label{eq:rangefinal}
\frac{\omega_{\rm B}}{a_0}<\omega<a_0\omega_{\rm B}\;.
\end{equation}

Models of the absorption process based on Eq.~\eqref{eq:gammaevo} \citep[][]{Lyubarsky2018, Beloborodov2024, Beloborodov2025} are problematic. Eq.~\eqref{eq:gammaevo}, which should describe the heating of the plasma caused by synchrotron absorption, was derived by considering the motion of a test particle in a monochromatic vacuum wave. We showed that the wave could be absorbed when $\omega_{\rm B}/a_0 < \omega < a_0\omega_{\rm B}$. This is exactly the regime in which, in general, the wave cannot be modeled as a monochromatic vacuum wave because the wave equation is strongly non-linear \citep[][]{Sobacchi+2024b}. Then, the assumptions used to derive Eq.~\eqref{eq:gammaevo} could break down exactly when Eq.~\eqref{eq:gammaevo} should be applied.

When $\omega_{\rm B}/a_0<\omega<a_0\omega_{\rm B}$, the wave-plasma interaction can be studied using fully kinetic simulations. Chen et al.~\citep[][]{Chen+2022} showed that the pulse profile can be significantly distorted, and each wavelength can steepen into a shock. Analytical descriptions of this process are based on the equations of magnetohydrodynamics \citep[][]{Lyubarsky2003, Beloborodov2024}. Eq.~\eqref{eq:gammaevo} does not describe the heating of the plasma because the heating occurs primarily at shocks. Simulations can be used to test whether Eq.~\eqref{eq:gammaevo} is applicable to some special cases, as suggested by ref.~\citep[][]{Beloborodov2024}.

\section{Discussion}
\label{sec:discussion}

We discussed synchrotron absorption of a short electromagnetic pulse that propagates in a cold magnetized pair plasma. We assumed that the particles are at rest before being illuminated by the pulse. We showed that synchrotron absorption is activated when $\tilde{\omega}\sim \tilde{\omega}_{\rm B}/\max[1,a]$, where $\tilde{\omega}$ is the frequency of the wave in the proper frame of the illuminated fluid, $\tilde{\omega}_{\rm B}$ is the cyclotron frequency in the fluid frame, and $a$ is the local strength parameter of the pulse. When $a>1$, the condition $\tilde{\omega}\sim\tilde{\omega}_{\rm B}/\max[1,a]$ is equivalent to the cyclotron resonance for a particle of effective mass $am$.

The resonance condition (i.e.~$\tilde{\omega}\sim\tilde{\omega}_{\rm B}/\max[1,a]$ in the fluid frame) can be satisfied when $\omega_{\rm B}/a_0 < \omega< a_0\omega_{\rm B}$, where $\omega$ and $\omega_{\rm B}$ respectively are the wave frequency and the cyclotron frequency in the lab frame (where the particles are at rest before being illuminated), and $a_0>1$ is the maximum strength parameter of the pulse.\footnote{We emphasize that the cyclotron frequency is defined as $\omega_{\rm B}=eB_{\rm bg,0}/mc$, where $B_{\rm bg,0}$ is the background magnetic field in the lab frame outside the pulse. The background magnetic field within the pulse can be larger than outside because the illuminated plasma can be compressed (as discussed in Sec.~\ref{sec:pulse}, we have $B_{\rm bg}\sim\gamma_{\rm bulk}^2 B_{\rm bg,0}$).} The pulse could be absorbed when $\omega_{\rm B}/a_0 < \omega< a_0\omega_{\rm B}$. In this regime, the propagation of electromagnetic waves in a plasma can be very different with respect to the propagation in vacuum because the wave equation is strongly non-linear \citep[][]{Sobacchi+2024b}. For example, each wavelength could steepen into shocks \citep[][]{Lyubarsky2003, Chen+2022, Beloborodov2024}. Fully kinetic simulations can be used to test whether particles can be heated stochastically due to synchrotron absorption, as found by studying the motion of a test particle in the field of a vacuum wave.

Our results can be important for constraining emission models of fast radio bursts (FRBs). In the magnetosphere and wind of magnetars, the most likely progenitors of FRBs, the condition $\omega_{\rm B}/a_0<\omega<a_0\omega_{\rm B}$ corresponds to distances $10^8{\rm\; cm}<R<10^{12}{\rm\; cm}$ from the star \citep[][]{Sobacchi+2024b}. As discussed in ref.~\citep[][]{Sobacchi+2024b}, FRBs may avoid absorption at these distances by propagating on top of a high-amplitude fast magnetosonic pulse, as in the reconnection-driven model of FRB emission \citep[][]{Lyubarsky2020, Mahlmann+2022}. Other mechanisms can prevent FRBs from escaping the magnetosphere and wind of magnetars. At small distances from the star ($R<10^8{\rm\; cm}$), one should consider the non-linear decay of fast magnetosonic waves into Alfv\'en waves \citep[][]{GolbraikhLyubarsky2023}. At large distances ($R>10^{12}{\rm\; cm}$), one should consider the induced Compton scattering \citep[][]{Lyubarsky2019, Nishiura2025}, the filamentation instability \citep[][]{Sobacchi+2025}, and the stimulated Brillouin scattering. The presence of protons in the plasma around the magnetar can also significantly affect the propagation of FRBs \citep[][]{Sobacchi+2024a}.

Relativistic perpendicular shocks produce a strong electromagnetic precursor wave that propagates into the upstream plasma and could be observed as an FRB \citep[][]{Lyubarsky2014, Beloborodov2017, Metzger+2019, Beloborodov2020, Sironi+2021, Iwamoto+2024, VanthieghemLevinson2025}. Our results imply that FRBs are not absorbed when produced through the shock maser mechanism. The properties of the precursor wave are extensively studied using fully kinetic simulations \citep[][]{Langdon+1988, Gallant+1992, Iwamoto+2017, Iwamoto+2018, PlotnikovSironi2019, Margalit+2020, Sironi+2021}. In the reference frame of the upstream plasma, the frequency and the strength parameter of the precursor wave are given by $\omega\sim 3\;\Gamma_{\rm sh}\omega_{\rm P}$ and $a_0\sim 3\times 10^{-2}\Gamma_{\rm sh}/\sqrt{1+\sigma}$, where $\Gamma_{\rm sh}$ is the Lorentz factor of the shock and $\sigma=\omega_{\rm B}^2/\omega_{\rm P}^2$ is the magnetization of the plasma \citep[][]{Iwamoto+2017, PlotnikovSironi2019, Margalit+2020, Sironi+2021}. Then, we have $\omega\sim 10^2 a_0\max[\omega_{\rm B},\omega_{\rm P}]\gg a_0\omega_{\rm B}$. Synchrotron absorption does not occur because the frequency of the precursor wave is too high with respect to the frequency range given by Eq.~\eqref{eq:rangefinal}.

\begin{acknowledgements}
This work was supported by a Rita Levi Montalcini fellowship. We thank the anonymous referees for their constructive comments. We acknowledge insightful discussions with Masanori Iwamoto, Amir Levinson, Yuri Lyubarsky, Lorenzo Sironi, and Arno Vanthieghem.
\end{acknowledgements}

\bibliography{2d}

\end{document}